\documentclass{article}
\pdfoutput=1
\usepackage{arxiv}
\usepackage[utf8]{inputenc} 
\usepackage[T1]{fontenc}    
\usepackage{hyperref}       
\usepackage{url}            
\usepackage{booktabs}       
\usepackage{amsfonts}       
\usepackage{nicefrac}       
\usepackage{microtype}      
\usepackage{graphicx}
\usepackage{natbib}
\usepackage{doi}

\usepackage{caption}
\usepackage{subcaption}

\title{A user study of visualisations of spatio-temporal eye tracking data}

\date{\today}
\author{Marcel Claus\\
	Zuyd University of Applied Sciences\\
	Nieuw Eyckholt 300\\
	Heerlen, 6419 DJ, The Netherlands \\
	\texttt{marcel.claus@outlook.com} \\
	\And
	Frouke Hermens
 \thanks{Corresponding author: Dr Frouke Hermens, Open University, PO Box 2960, 6401 DL Heerlen, the Netherlands. E-mail: frouke.hermens@ou.nl. The authors declare that they have no competing interests. The anonymised data from the survey and the original eye tracking study can be downloaded from: \href{https://osf.io/dz69n/}{https://osf.io/dz69n/}.}\\
	Department of Computer Science\\
	Open University of the Netherlands\\
	Valkenburgerweg 177 \\
 Heerlen, 6419 AT, The Netherlands\\
	\texttt{frouke.hermens@ou.nl} \\
	\AND
	 Stefano Bromuri \\
Department of Computer Science\\
	Open University of the Netherlands\\
	Valkenburgerweg 177 \\
 Heerlen, 6419 AT, The Netherlands\\
	\texttt{stefano.bromuri@ou.nl} \\
}

\renewcommand{\shorttitle}{Visualising spatio-temporal aspects of eye tracking data}

\hypersetup{
pdftitle={Visualising spatio-temporal aspects of eye tracking data},
pdfsubject={cs.HC, cs.OH},
pdfauthor={Marcel Claus, Frouk Hermens, Stefano Bromuri},
pdfkeywords={data visualisation, eye tracking, spatio-temporal data, user study},
}

\begin{document}
\maketitle

\begin{abstract}
Eye movements have a spatial (where people look), but also a temporal (when people look) component. Various types of visualizations have been proposed that take this spatio-temporal nature of the data into account, but it is unclear how well each one can be interpreted and whether such interpretation depends on the question asked about the data or the nature of the data-set that is being visualised. In this study, four spatio-temporal visualization techniques for eye movements (chord diagram, scanpath, scarfplot, space-time cube) were compared in a user study. Participants (N = 25) answered three questions (what region first, what region most, which regions most between) about each visualization, which was based on two types of data-sets (eye movements towards adverts, eye movements towards pairs of gambles). Accuracy of the answers depended on a combination of the data-set, the question that needed to answered, and the type of visualization. For most questions, the scanpath, which did not use area of interest (AOI) information, resulted in lower accuracy than the other graphs. This suggests that AOIs improve the information conveyed by graphs. No effects of experience with reading graphs (for work or not for work) or education on accuracy of the answer was found. The results therefore suggest that there is no single best visualisation of the spatio-temporal aspects of eye movements. When visualising eye movement data, a user study may therefore be beneficial to determine the optimal visualization of the data-set and research question at hand.	
\end{abstract}
\keywords{data visualisation \and eye tracking \and spatio-temporal data \and user study}

\section{Introduction}

Humans make an eye movement around 2 to 5 times per second in order to refocus the sensitive part of the eye (the fovea) to the area of the scene that is of interest to the current task \citep{rayner1978eye,rayner1998eye}. Recent developments in eye tracking technology allow for the accurate recording of these eye movements not only while reading text on a computer screen \citep{rayner2005eye}, but also in real-world tasks, such as driving \citep{land1996relations} and making a cup of tea \citep{land2001ways}.

In order to understand the pattern of eye movements recorded by these eye tracking devices, it is important to create visualizations of the recorded data that are intuitive and easy to understand \citep{blascheck2017visualization}. Attempts to visualize eye movements have often focused on where people look \citep[e.g., by the use of heatmaps,][]{bojko2009informative,menges2020visualization} \citep[or the related bee swarm,][]{burch2021power,bakardzhiev2021web} but on their own often fail to capture when, in what sequence and how long people look at different parts of the scene. When attempts are made to represent both time and space \citep[e.g., using scanpaths,][]{noton1971scanpaths_a}, the issue often arises that plots of data of all participants often look cluttered \citep{peysakhovich2017scanpath,rodrigues2018multiscale}.
 
While various types of data visualisations have been proposed for eye tracking data \citep[for an overview, see,][]{blascheck2017visualization}, only few studies have examined the adequacy of these visualisations in a user study. In studies in which users were asked about visualisations of eye movements, questions were often about their subjective experience about the graphs \citep{blascheck2017visualization}, rather than asking users to solve questions about the data with the visualisations \citep[for example,][asked specific questions, but did not report the results]{menges2020visualization}. An exception is the survey by \cite{burch2021power}, where participants were asked to solve four different tasks for a single data-set, allowing them to choose how many of the visualisations they used for the task. This study showed that the chosen combination of visualisations depended on the task.

Studies do not typically systematically vary the data-set that is used to create the visualisations \citep[e.g.,][]{eraslan2016eye,menges2020visualization}. Different data-sets, however, can lead to different levels of clutter in visualisations. If the eye movements are focused on a small number of aspects of a visual scene, higher levels of clutter in visualisations can be expected compared to when eye movements cover most of the visual scene. One study that compared multiple data-sets (eye tracking of two static stimuli and one dynamic stimulus) is the user study of a newly introduced visualisation method \citep[gaze stripes,][]{kurzhals2015gaze}, but the authors used a qualitative analysis of the answers, and focused on a single visualisation, meaning no quantitative comparison was made between different visualisations. 
 
The present study therefore investigates whether the optimal visualisation may depend on the combination of task and data-set by asking participants to answer questions about eye movement data presented in four different spatio-temporal visualisations (chord diagram, scanpath, scarfplot, space-time cube). To examine the effect of the question about the data, three different questions were asked: (1) where did people look most? (2) where did people look first? (3) which two regions did people look between most often?. To determine the effects of the type of data, two types of data were used: One in which people looked at adverts (where we expect eye movements to cover the entire images) and one in which people looked at numbers representing two gambles (where we expect eye movements to be focused on the four numbers on the screen). We test the hypothesis that the accuracy on the questions will depend on the combination of the type of visualization, the question asked and the type of eye movement data used for the visualisation.

\section{Related work}

An extensive overview of proposed visualisations for eye tracking data is provided by  \cite{blascheck2017visualization} \citep[see also,][]{blascheck2014state}. They distinguish between point-based and areas-of-interest (AOI) based visualisations (or visualisations that are a combination), referring to whether the graphs show the areas of interest, such as in a scarfplot, or the individual fixations, such as in a scanpath. They also indicate which of the visualisations provide temporal only, spatial only, or spatio-temporal information. For the current investigation, those showing spatio-temporal patterns are of most interest. There are also distinctions between whether data of individual participants or multiple participants are shown, whether visualisations are static or animated (or interactive), whether 2D or 3D is used and whether graphs incorporate the original stimulus or not \citep{blascheck2017visualization}. 

An important distinction, that  \cite{blascheck2017visualization} also make, is whether eye movements were recorded for static images, animations, interactions (e.g., on a computer screen) or in real-world settings. Some visualisations only provide insight when what participants viewed, is constant, not only between participants (e.g., for animations) but also within participants. An example is the scanpath \citep{noton1971scanpaths,noton1971scanpaths_a}, where fixations are shown directly superimposed onto the image. Animated scanpaths are needed when this image changes over time. It gets more complicated when each participant had their own visual input, such as during interactions with a computer screen or in real-world eye tracking \citep[e.g.,][]{land1996relations}. In the present investigation, we only consider the case of static images.

\subsection{Scanpaths}

One of the visualisations that we consider in our comparison is the scanpath \citep{noton1971scanpaths_a,noton1971scanpaths}, illustrated in Figure~\ref{fig:visualisations}B. It consists of the original image, with fixations and saccades superimposed, where fixations are typically represented by circles or dots and saccades by lines or arrows. Fixations are often numbered to indicate their sequence. Sometimes their duration is represented by the size of the circles. As can be seen in Figure~\ref{fig:visualisations}B, where data of only five participants are shown, the image can quickly become cluttered when the number of participants or the recording time increases. 

\cite{blascheck2017visualization} discusses approaches from the literature to reduce the clutter \citep{hembrooke2006averaging,chen2013combining,hurter2013bundled}, but also indicates that the question of scanpath similarly (needed for adequate clustering) has not been fully answered yet. Fixations in such approaches are typically clustered on the basis of their spatial distance \citep{peysakhovich2017scanpath}, but this does not take into account whether they were directed towards the same objects on the screen. In Figure~\ref{fig:visualisations}B, for example, fixations could be spatially close, but either directed towards a product, or the adjacent price-tag of the product. By pooling such fixations, subtle differences between where people look may be discarded. An alternative approach was presented by \cite{menges2020visualization} in which a heatmap was combined with a scanpath. This approach, however, requires the visualisation to be interactive, and therefore falls outside the scope of the present study.

\subsection{Space-time cube}

Standard graphs can present two spatial dimensions (using the horizontal and vertical components), which can be a limitation when presenting spatio-temporal data such as from eye movements, which have three components (horizontal, vertical, time). One option is to plot the horizontal component as a function of time, as well as the vertical component as a function of time, but this makes it difficult to relate the two components to the image that was presented to viewers.

It may therefore be beneficial to make use of 3D graphs, that allow for the simultaneous presentation of the horizontal, vertical and time component. 3D visualisations are often discouraged, as the perspective may make estimation of the size of elements in the graph difficult, in addition to possible occlusion of elements by other elements \citep[e.g., a bar hiding behind another bar,][]{elmqvist2005balloonprobe}. Such issues may become less problematic when the user can rotate the graph and look at it from multiple angles \citep{elmqvist2008taxonomy}

One type of 3D visualisation that has been proposed for eye movement data is the space-time cube \citep{kurzhals2013space,kurzhals2015task}, illustrated in Figure~\ref{fig:visualisations}C. At the bottom of the graph \citep[in the original version of the spac-time cube, time is shown from left to right,][]{kurzhals2013space}, the original image, or the AOI image is shown. Traces start from this image and grow as viewing time increases. Sections of these traces that belong to a particular AOI are shown in the colour corresponding to that AOI. The graph can be interactive (allowing participants to rotate it), or dynamic \citep[moving the original image along the time-axis,][]{kurzhals2013space}. A similar type of graph was presented by \cite{menges2020visualization}, but without an interactive component, for which it was criticised by the users that they surveyed.

\subsection{Scarfplot}

While the space-time cube makes use of AOIs, it could be presented without the AOI information and still provide insight into the data \citep{menges2020visualization}. There are, however, visualisations that plot eye tracking data based on the AOI information. A powerful method for comparing sequences of eye movements is the scarfplot \citep{burch2021power,bakardzhiev2021web,richardson2005looking,kurzhals2015task}, where the horizontal axis shows time and the vertical axis the various AOIs (see Figure~\ref{fig:visualisations}A). The scarfplot makes it easy to compare patterns of eye movements between participants. Participants can be ordered on the basis of the similarity in their patterns, so clusters of eye movement patterns are easier to identify \citep{kurzhals2015gaze}. Scarfplots can leave spaces during saccades and blinks or merge these intervals \citep{blascheck2017visualization} with the preceding fixation (as in Figure~\ref{fig:visualisations}A).

While scarfplots allow for the comparison of many participants, they become less effective when the number of AOIs is large \citep{yang2018alpscarf}. To address this issue, \cite{yang2018alpscarf} suggested the use of mountains and valleys denoting the various sub-AOIs of each AOI. For our data, we use a small number of AOIs, and therefore can rely on standard scarfplots.

\subsection{Chord diagram}

The scarfplot, scanpath and space-time cube all focus on where participants looked over time, but may be less suited to study how often participants shift their gaze between regions. Several visualisations have been proposed to better show such transitions between regions. 

An often used plot is the transition matrix \citep[e.g.,][]{goldberg1999computer,krejtz2013effects}, where AOIs are shown along the horizontal and vertical axis and the colour code indicates the frequency of eye movements between the various regions. A similar graph is the transition graph, where AOIs are shown as coloured sections of a ring, with arrows indicating the frequency of eye movements between these AOIs \citep{blascheck2013circular}.

We make use of a variation of radial graphs \citep{goldberg2011eye}, namely the chord diagram (Figure~\ref{fig:visualisations}D). Chord diagrams are commonly used to show relations \citep{rees2020interaction} and have been proposed in the context of eye tracking \citep{wang2021eye}. The chord diagram shows saccades as individual curved lines originating from the section of a ring representing the corresponding starting AOI. Chord diagrams are symmetrical in that they do not provide arrowheads or other indications of what AOI was the starting or the end point.

\subsection{Alternative approaches}

We restricted our user study to four possible visualisations to avoid over-asking our participants. Several other approaches, however, have been suggested. An extensive overview can be found in \cite{blascheck2017visualization}, but we would like to mention a few of these here.

One spatio-temporal visualisation of eye tracking data is the time plot, where discrete AOIs are plotted on the vertical axis and time is shown on the vertical axis, with the size of the symbols indicating the duration of the fixations \citep{raiha2005static}. The time plot will become cluttered quickly if data from more than one participant is presented.

AOI rivers \citep{burch2013aoi} are similar to time plots, but with the possibility to represent data from multiple participants (wider bands indicating more participants `flowing' from one AOI to another). AOI rivers resemble Sankey diagrams \citep{riehmann2005interactive}. They allow for visualising longer sequences than for example the transition graph and chord diagram in which only transitions between two AOIs can be shown. While \cite{burch2013aoi} provide the algorithm to generate AOI rivers, implementation may be fairly complicated. As indicated by \cite{burch2013aoi}, AOI rivers work best for a limited number of AOIs.

A further visualisation is the gaze stripe \citep{kurzhals2015gaze,kurzhals2015task}, which uses small icons of showing the region of the image that was fixated over time (presented on the horizontal axis). The advantage of gaze stripes is that no areas of interest need to be defined, but that it is still possible to compare eye movement patterns of different participants. A disadvantage of the plot is that if regions in the image are highly similar, but further apart it will be difficult to tell which patterns overlap in terms of meaning only, or also in spatial location. A further difficulty is implementing the algorithm to create the visualisation. It is unclear how well gaze stripes work for data-sets of a large number of participants (e.g., more than 30).

\section{Methods}

\subsection{Participants}

A total of 25 participants took part in the study. An opportunity sample was used. As a consequence, most of the participants were staff and students at Zuyd University of Applied Sciences (the Netherlands). Participants all provided informed consent for their participation in the study that was approved by the ethics committee of the Open University of the Netherlands.

\subsection{Apparatus}

The experiment was conducted on a MacBook Pro running Monterey 12.3 with a M1 Pro chip (8-Core CPU, 14-Core GPU), 16GB Unified Memory, 512GB SSD Storage and a 14-inch Liquid Retina XDR display. The screen was set to a 1024 by 768 pixels resolution. Responses were collected using the keyboard of the laptop.

\subsection{Stimuli}

\subsubsection{Original data-set}

The stimuli in the present user study were based on data from an unpublished eye tracking study on the role of advance price information on subsequent choices between gambles, conducted in the context of a dissertation project at the University of Leuven (Belgium). Ethics approval for this eye tracking study was provided by the psychology ethics committee of the University of Leuven.

The twenty-seven participants of this study were shown with a total of sixty-two adverts from various shops (see Figure~\ref{fig:images} for examples) that either had an overall high or an overall low price level, each presented for 3 seconds. The adverts were followed by the presentation of two gambles in the form of `a X\% chance of Y Euro' (but presented in Dutch), presented across two lines. Participants indicated, with a button press, which of the two gambles they preferred, after which the computer played the corresponding gamble and provided feedback about the outcome and the total amount won. 

Eye movements in this study were recorded with an Eyelink 1000 (SR Research, ON, Canada) desk-mount eye tracker (monocular recording of the movements of the left eye at 1000Hz). The experiment produced two types of eye tracking data from two intervals of each trial: Eye movements when viewing an advert (for a fixed amount of time) and eye movements when deciding between two gambles (variable measurement time between participants, dependent on how quickly participants responded). The resulting data from these two sections of each trial differ in the duration of the recording interval (3 seconds for the adverts, variable for the gambles). They are also likely to vary in the distribution of spatial locations of the fixations (four positions for the gambles, many positions for the adverts). 

\subsubsection{Visualisation types}

\begin{figure*}[h]
\includegraphics[width=0.95\textwidth]{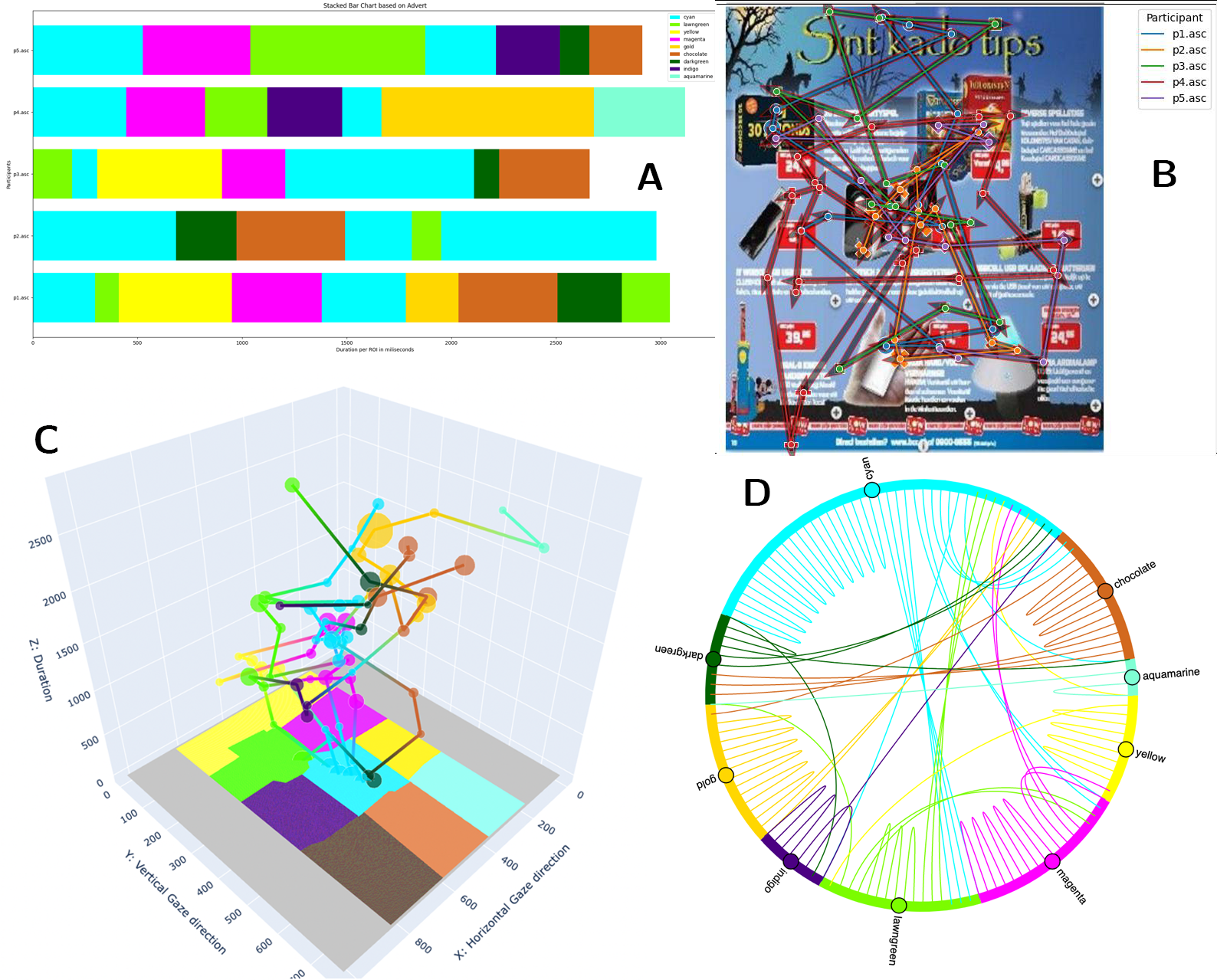}
\caption{Examples of the four types of visualisations used in this study. A) Scarfplot, B) scanpath, C) space-time cube, D) chord diagram. All are based on data from the same five participants of the eye tracking data-set.}
\label{fig:visualisations}
\end{figure*}

Participants in our user study saw images of four types of visualisations: (1) a scarfplot, (2) a scanpath, (3) a space-time cube, and (4) a chord diagram (Figure~\ref{fig:visualisations}). Three of these visualisations make use of the areas of interest associated with the images (the scarfplot, the space-time cube and the chord diagram), whereas the scanpath plots the eye movements directly superimposed onto the raw image. 

For each type of visualisation, data was presented of the same five participants of the original eye tracking study. The choice was made to use only five of the original participants, because inspection of the graphs for all original 27 participants indicated that several of the graphs became too cluttered for interpretation. This means that we made the assumption that when applying the different visualisations, users would select data of around five participants at each time. 

\begin{figure*}[h]
\includegraphics[width=0.95\textwidth]{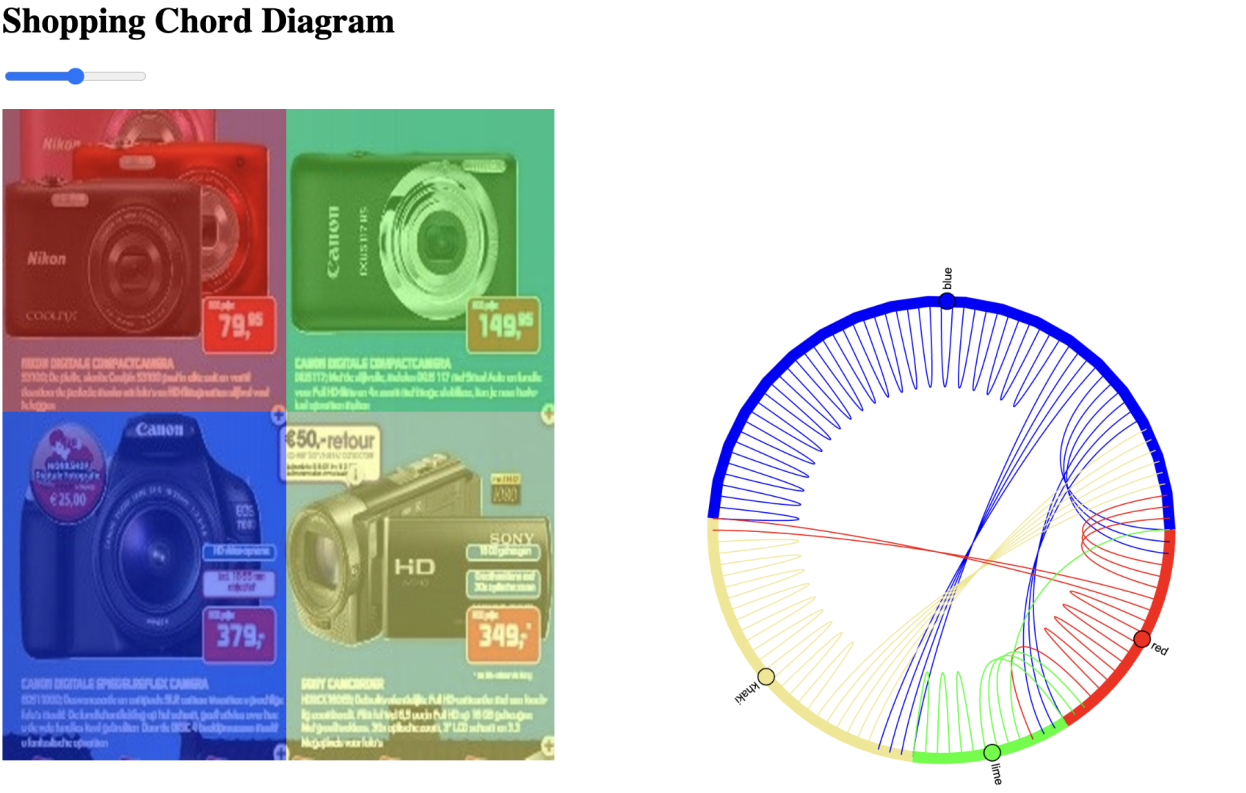}
\caption{Example of how visualisations were presented to participants. On the left, participants saw a superposition of the original image and the areas of interest (AOI) image. They could use the slider above the image to change the transparency of the AOI image, so that they could vary between seeing just the original image, just the AOI image, or a mixture of the two images. On the right, the visualisation of the data was shown (a chord diagram in this case). The task of participants was to indicate the colour corresponding to the answer to the question (e.g., `Which region was looked at most?', where the correct answer in this instance would be `blue'. In the case of a scanpath (which was the only visualisation that did not directly use AOIs) the same combination of the original and AOI image was shown and participants had to use this image to provide their answer (e.g., decide that the lower left camera and its price was looked at most, corresponding to the blue region).}
\label{fig:AOIs}
\end{figure*}

\begin{figure*}[h!]
\begin{subfigure}{.99\textwidth}
\caption{Advert images}
    \includegraphics[width=0.95\textwidth]{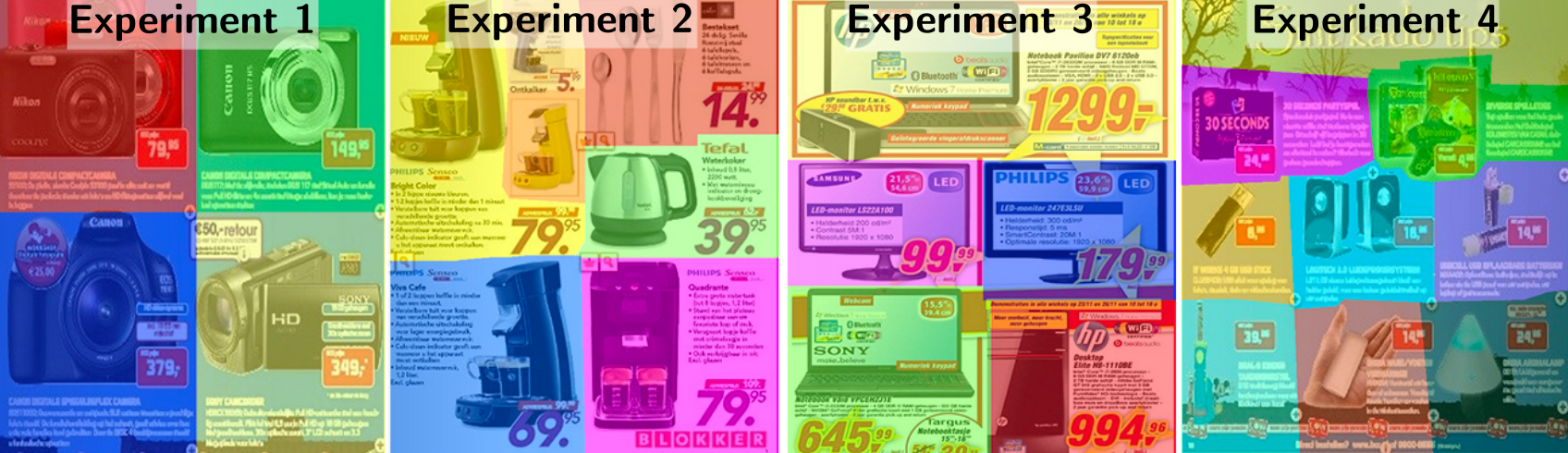}   
\end{subfigure}
\begin{subfigure}{.99\textwidth}
\caption{Gambling images}
    \includegraphics[width=0.95\textwidth]{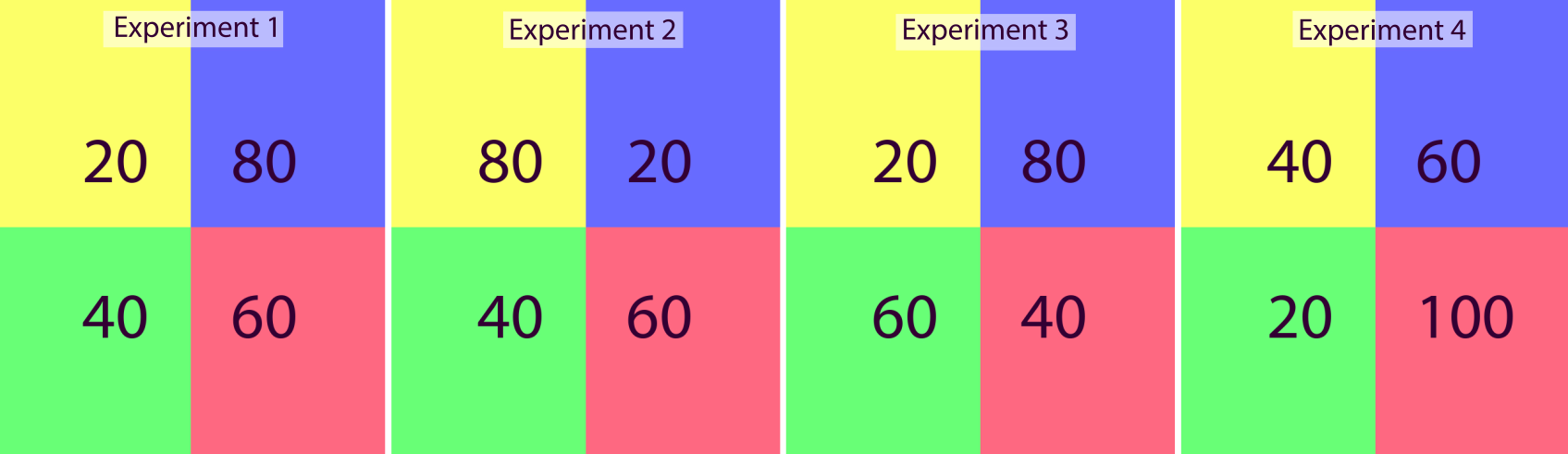}     
\end{subfigure}
\caption{Illustrations of eight images for which eye tracking data were visualised. These images were from individual trials in the original eye tracking experiment, in which participants first saw the (a) advert, followed by (b) a choice between two gambles. `Experiment' in these illustrations refers to the different subgroups of participants in our user study (who performed the same tasks, but saw a different image for each task). For the adverts, we expected eye tracking data to cover a larger part of the image than for the gambling images, where we expected observers to mostly look at the four numbers on the screen. Note also that the number of colours for the gambling images was always equal to four, whereas for the adverts, a larger number of AOIs, and therefore colours could be present in the AOI image.}
\label{fig:images}
\end{figure*}

Images were presented superimposed on the associated regions of interests, as illustrated in Figure~\ref{fig:AOIs}. To reduce the effects of the particular layout of the original image on the visualisation (mostly important for the advert images, as the images with gambles were more uniform), four different images were used (one for each group of participants, see Figure~\ref{fig:images}). Where possible, analyses will pool the data across the different images, but in some of the analyses, data can only be interpreted per group of participants and are therefore shown as such.

Visualisations were created with Python, using the Pandas (data handling), Seaborn (scarfplot), Matplotlib (scanpath), Plotly (space-time cube) and Holoviews (chord diagram) packages. While the scarfplot and scanpaths were fully static, the space-time cube and the chord diagram provided some interaction (rotation in the case of the space-time cube, and selection for the chord diagram). AOIs for the different images (see Figure~\ref{fig:images}) were manually created using the Photoshop software packages.

\subsection{Design}

Participants, unknowingly to them, were assigned to one of four conditions, each with their unique order of the visualisations, order of the questions and particular images used to generate the visualisations (Figure~\ref{fig:images}). The order of the visualisations was varied systematically, so that each visualisation was presented equally often first, second, third and last. In total, participants answered 24 questions, which were a combination of 4 visualisations, 3 types of questions and 2 types of data.

\subsection{Procedure}

Participants were invited to the study by e-mail or in person. They all participated in person. After reading an information sheet and providing informed consent, they were shown the different visualisations one by one on a laptop screen. They answered the questions about these visualisations by entering their response into a text box in a separate questionnaire. One of the answers that participants could provide is that an answer could not be determined from the data visualisation. For one of the visualisations, the chord diagram and the question where people looked first, this was the correct answer. The final questions of the questionnaire asked participants for their highest attained education level, and the frequency with which they worked with graphs while at work (daily, weekly, monthly, less frequently) and away from work (e.g., in newspapers, blogs, social media).

\section{Results}

Answers can be scored in an all or none (i.e., the answer is correct or incorrect) way, or using partial scores (e.g., if for a combination of two regions, only one of the two regions is correct). We here provide results for the all-or-none scoring. Similar outcomes were obtained for a partial scoring (i.e., the conclusions from the analyses were the same).

\begin{figure*}[h!]
    \includegraphics[width=0.95\textwidth]{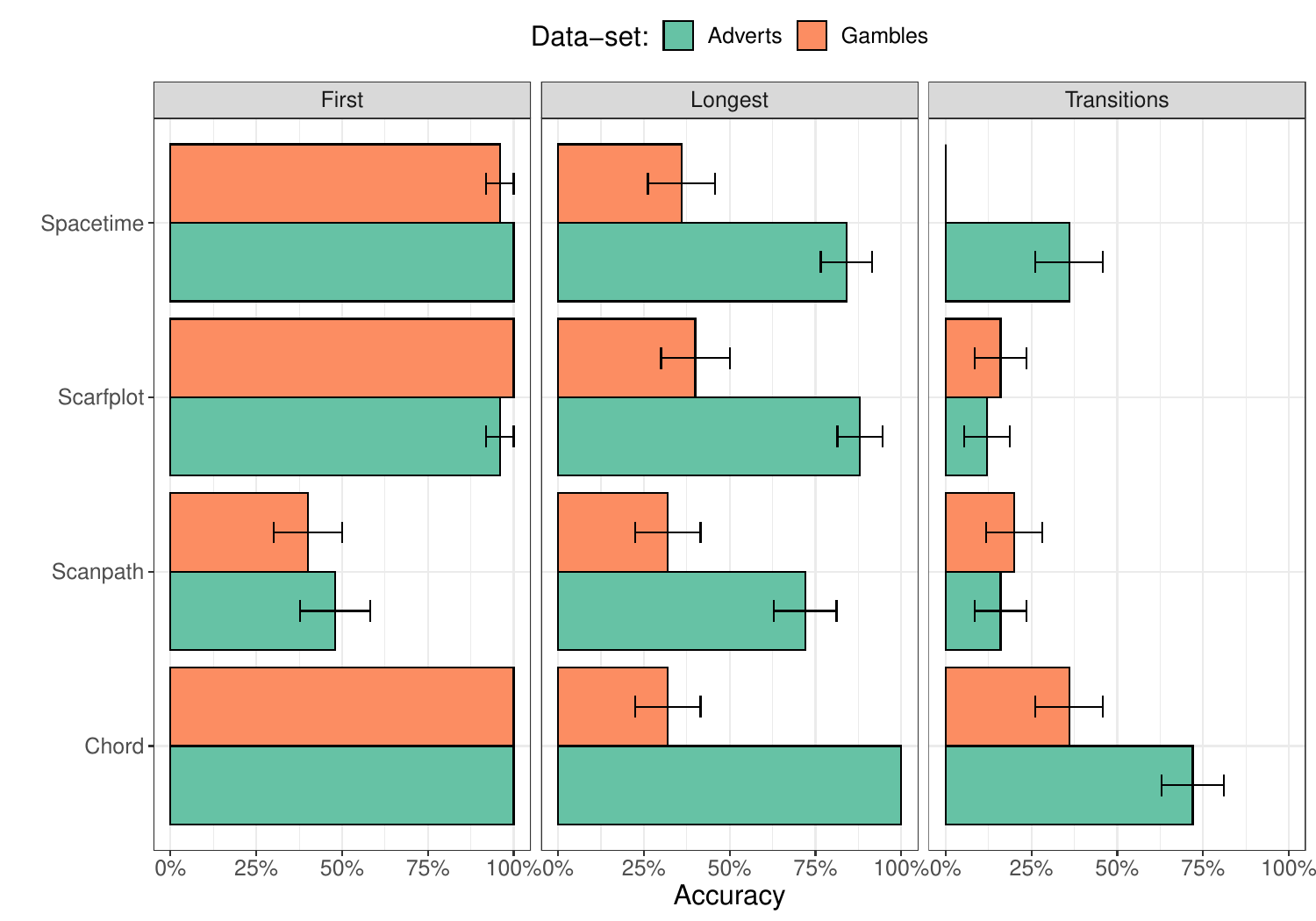} \caption{Accuracy of the responses for each combination of type of visualisation (vertical axis), data-set (different colours), question about the data (panels). Error bars show the standard error of the mean across participants.}
\label{fig:accuracy}
\end{figure*}

Figure~\ref{fig:accuracy} shows the percentage of correct answers per combination of visualisation, data-set, and question asked about the data. The plot suggests that ability to answer the questions correctly depends on the combination of data-set, visualisation and question asked, as hypothesized. Accuracy is particularly low for the question which two regions observers looked most often between (`transitions'). Accuracy is also low for the scanpath visualisation. Lower accuracy is found for visualisations of the eye movements during presentation of the gambles compared to presentation of the adverts, particularly when asked about which region was looked at longest. Note that accuracy is high for the combination of a chord diagram and which area was looked at first, because participants were always correct in stating that this information could not be determined from the visualisation.

A mixed effects logistic regression, comparing a full model with a three-way interaction (and two-way interactions and main effects) against a model with two-way interactions (and main effects) demonstrated a significant three-way interaction between type of visualisation, question, and data-set ($\chi^2(6)$ = 12.8, $p$ = 0.047), confirming the observation from visual inspection of the results that accuracy of the response depends on the combination of visualisation, question, and data-set. For the gambles data-set, we find a significant interaction between visualisation and question ($\chi^2(6)$ = 43.6, $p <$0.001), indicating that some visualisations are better suited to answer certain questions about this data-set. Also for adverts data-set, we find a significant interaction between visualisation and question ($\chi^2(6)$ = 17.8, $p=$ 0.0068).

Tests of the effect of visualisation within each combination of question and data-set show significant differences for `transitions' for the gambles data-set ($\chi^2(3)$ = 14.6, $p$ = 0.0022), `longest' for the adverts data-set ($\chi^2(3)$ = 11.2, $p$ = 0.011) and `transitions for the adverts data-set ($\chi^2(3)$ = 26.8, $p <$ 0.001), but not for `longest' for the gambles data-set ($\chi^2(3)$ = 0.48, $p$ = 0.92). For the question which region was looked at first, a statistical test could not be performed, because for at least two of the visualisations all responses were correct.

The main take-away from these results is that the ability to correctly answer questions about the data depends on the combination of visualisation, data-set and question.

\subsection{`Unknown' answers}

\begin{figure*}[h!]
    \includegraphics[width=0.95\textwidth]{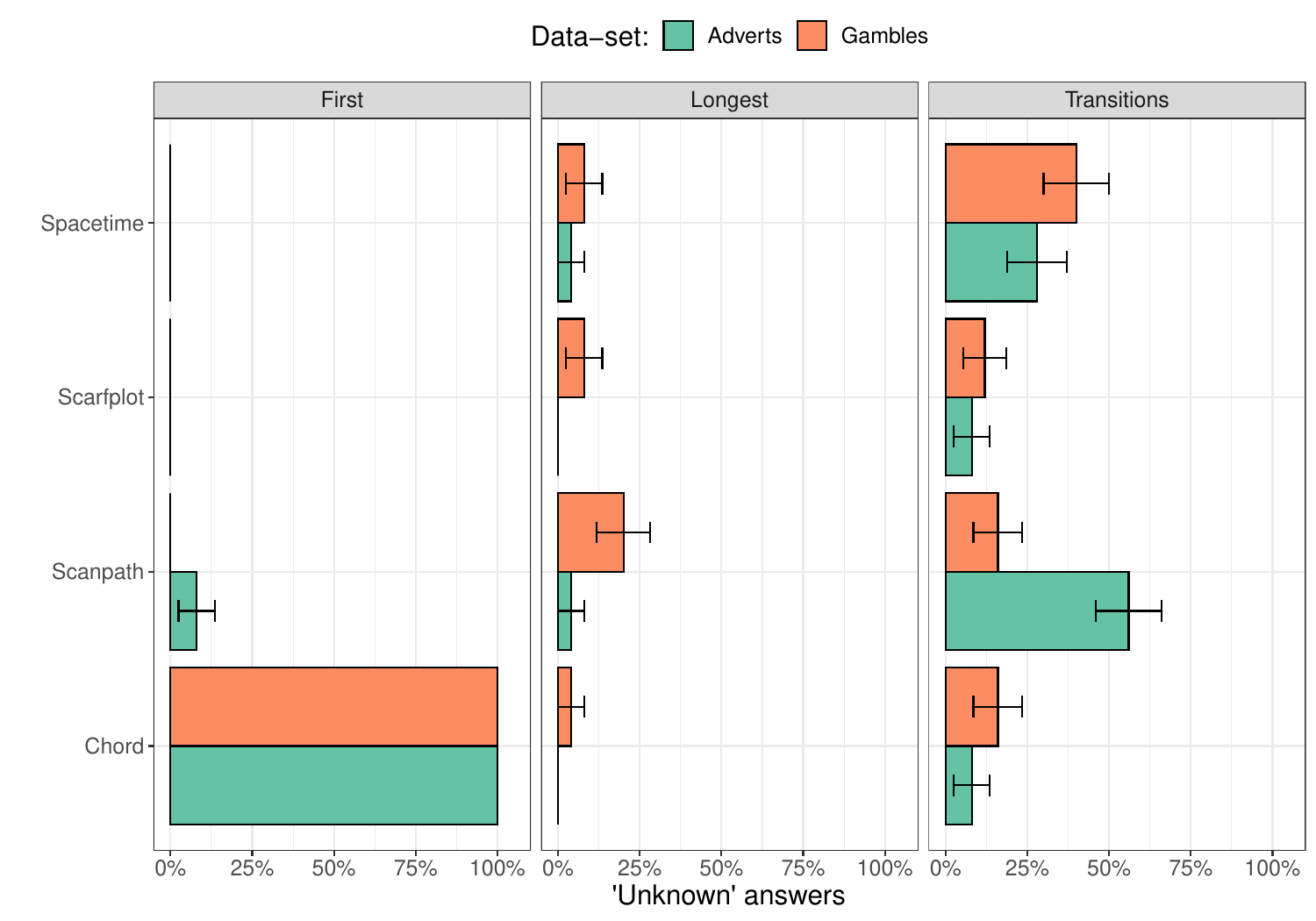} \caption{Percentage of `unknown' answers for each combination of type of visualisation (vertical axis), data-set (different colours), question about the data (panels). Error bars show the standard error of the mean across participants.}
\label{fig:unknown}
\end{figure*}

Participants sometimes indicated that the answer could not be derived from the plot. Figure~\ref{fig:unknown} provides the rates of these responses per combination of visualisation, question and data-set. For the combination of the chord diagram and the question about which region was looked at first by most participants, the correct answer was that the visualisation did not contain that information (`unknown'), indicated correctly by all participants. The estimation of the mixed effects models predicting whether or not participants responded with `unknown' failed to converge (meaning no reliable result could be obtained), so we here only present the average data with their confidence intervals (standard error of the mean).

Participants more frequency indicated that the answer could not be derived from the graph for the question about between which regions observers shifted their gaze most (`transitions') than for the other two questions (except for the combination of `first region' and `chord diagram', where the correct answer was `unknown'). Higher rates of `unknown' responses were found for the scanpath and the space-time cube. These data align with the accuracy data in suggesting that the scanpath was more difficult to interpret than the other visualisations.

\subsection{Specific images}

\begin{figure*}[h!]
    \includegraphics[width=0.95\textwidth]{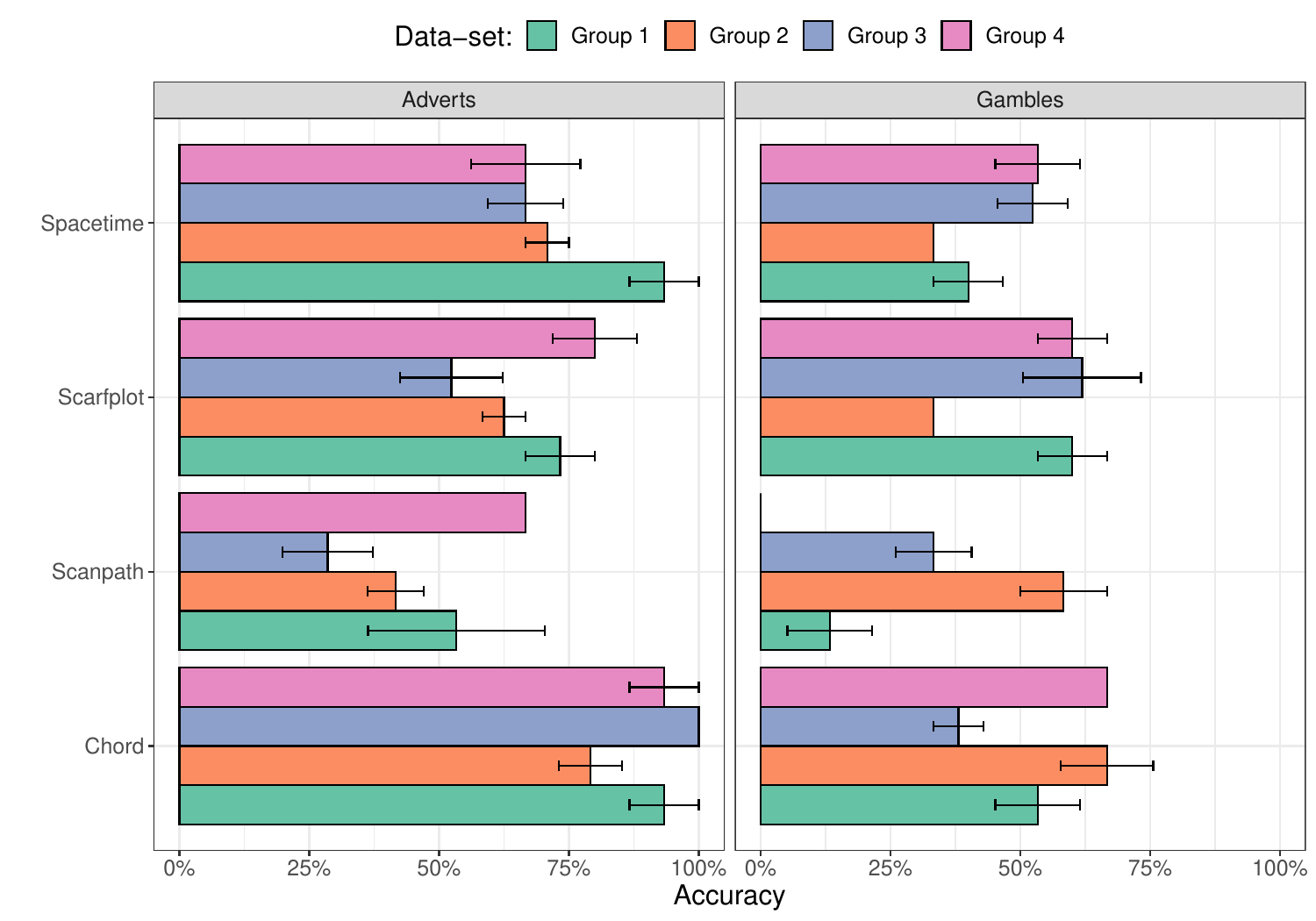} 
    \caption{Accuracy of the responses for each combination of type of visualisation (vertical axis), data-set (different panels), image that participants saw (colours - which could be an advert or a pair of gambles, together with the corresponding eye tracking data). Error bars show the standard error of the mean across participants.}
\label{fig:groups}
\end{figure*}

We varied the images (or gambles) for which participants interpreted the visualisations across four groups of participants. Figure~\ref{fig:groups} examines the effects of the image (and corresponding eye tracking data) on the overall accuracy (across the three questions) per combination of visualisation and type of data (viewing gambles or adverts). Figure~\ref{fig:groups} shows that accuracy depends on the specific image (or gamble) for which the data were shown, but that overall the worse performing visualisations (the scanpath in particular) had lower accuracy than better performing visualisations (the chord diagram in particular).

Interestingly, there is similar variation between subsets of participants (and therefore advert images or gambles and their corresponding eye tracking data) for the gambles and the adverts, even though the AOIs were much more homogeneous for the adverts (in shape, size and number). Group 4 had the largest number of AOIs, group 1 the smallest number, and groups 2 and 3 an intermediate number. The similarly between variations in accuracy between gambles (always four AOIs) and adverts (varying numbers of AOIs) suggests that the number of AOIs did not play an important role. Overall, these results indicate that it is important not to rely on a single image for which data are collected (because otherwise the results could depend strongly on the one image selected).

\subsection{Alternative responses}

For some of the images, the distinction between the top answer and second best answer was difficult to make. For example, the blue region could be fixated for an average of 700ms (top answer) and the red region for 690ms (second best answer). In such a situation the accuracy for the longest fixated region can be expected to be lower than when blue has 700ms (top answer) and the red region 350ms (second best answer). 

\begin{figure*}[h!]
\begin{subfigure}{.49\textwidth}
\caption{Adverts, first fixation}
    \includegraphics[width=0.95\textwidth]{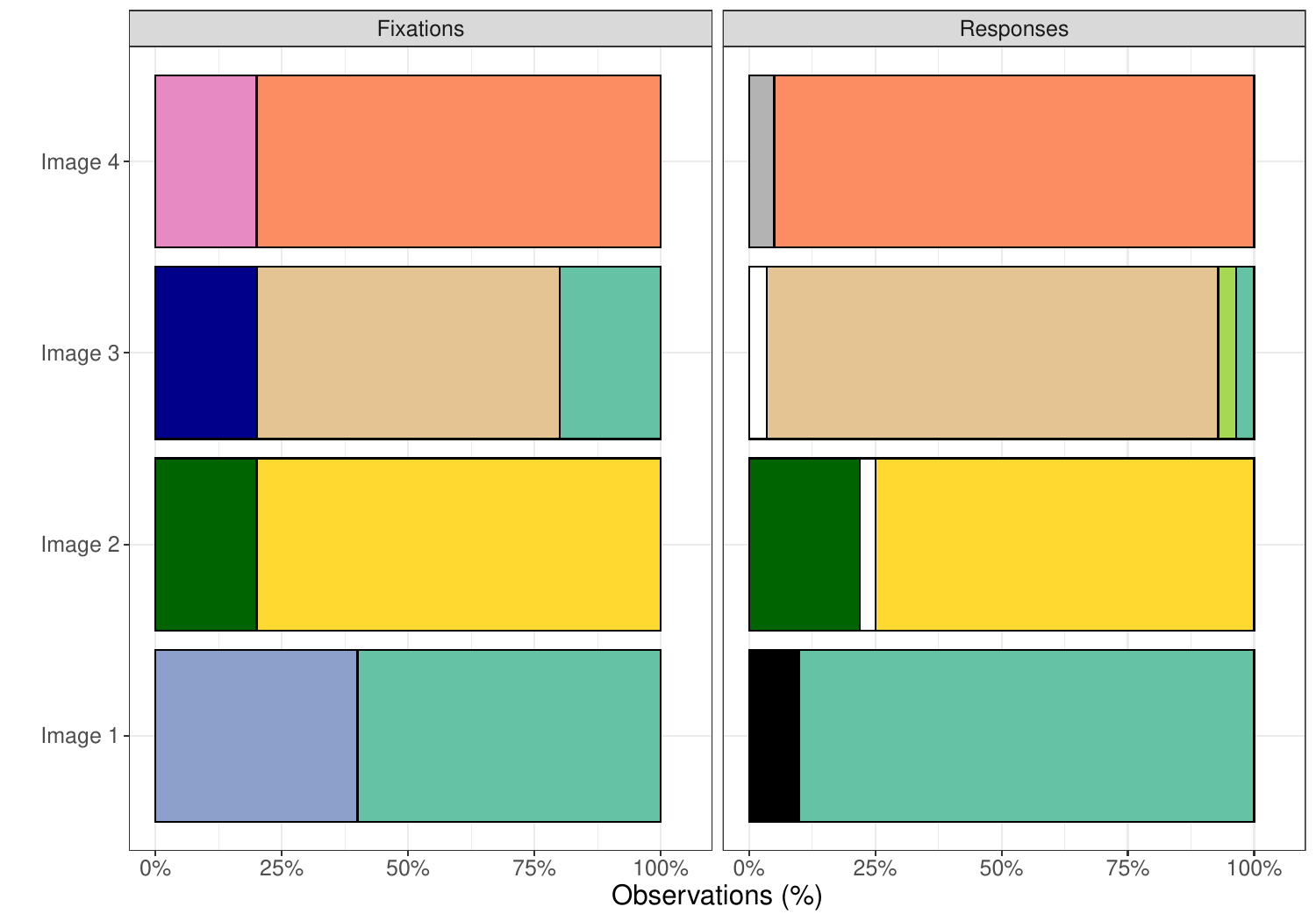}   
\end{subfigure}
\begin{subfigure}{.49\textwidth}
\caption{Adverts, longest dwell time}
    \includegraphics[width=0.95\textwidth]{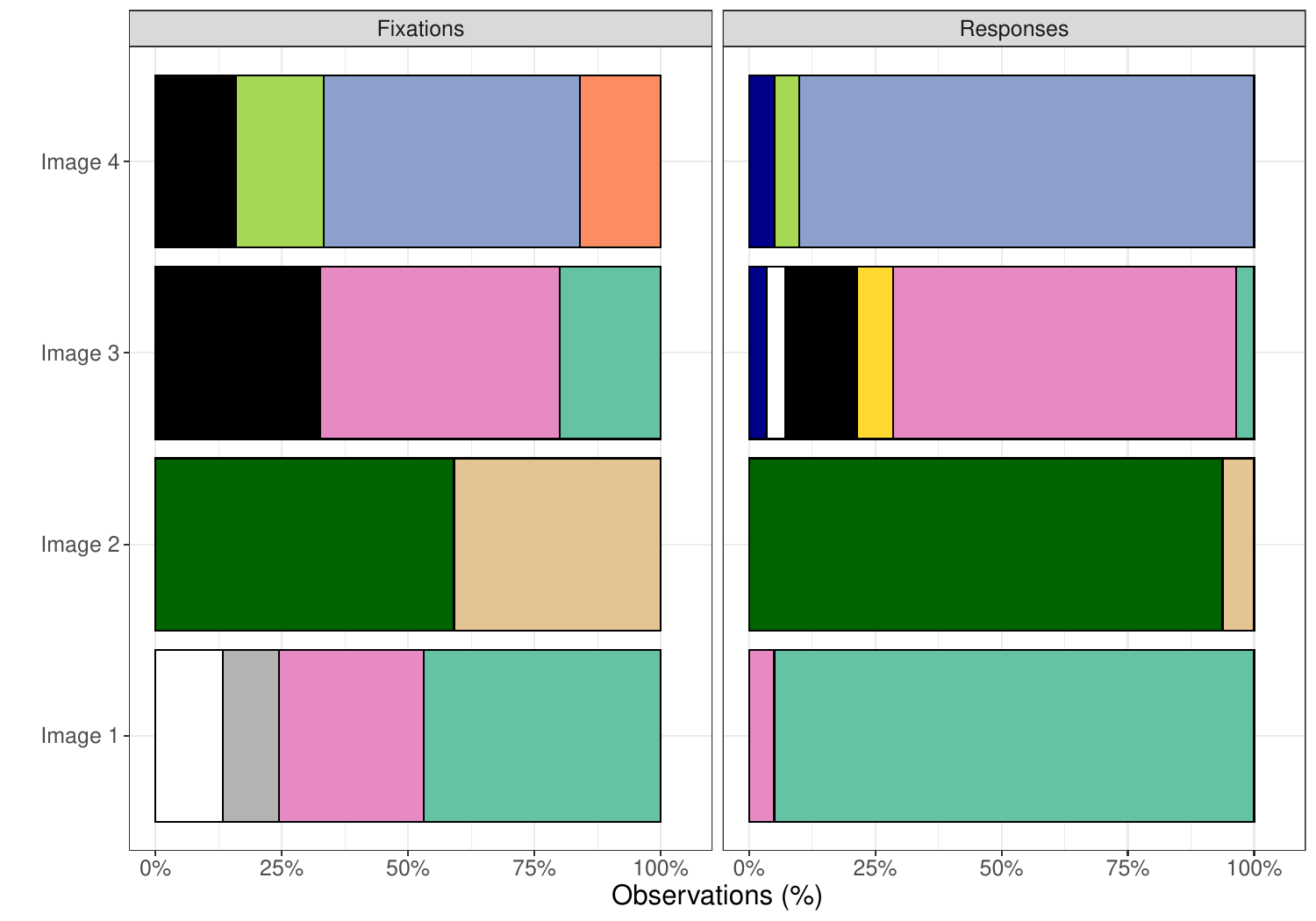}     
\end{subfigure}
\begin{subfigure}{.49\textwidth}
\caption{Gambles, first fixation}
    \includegraphics[width=0.95\textwidth]{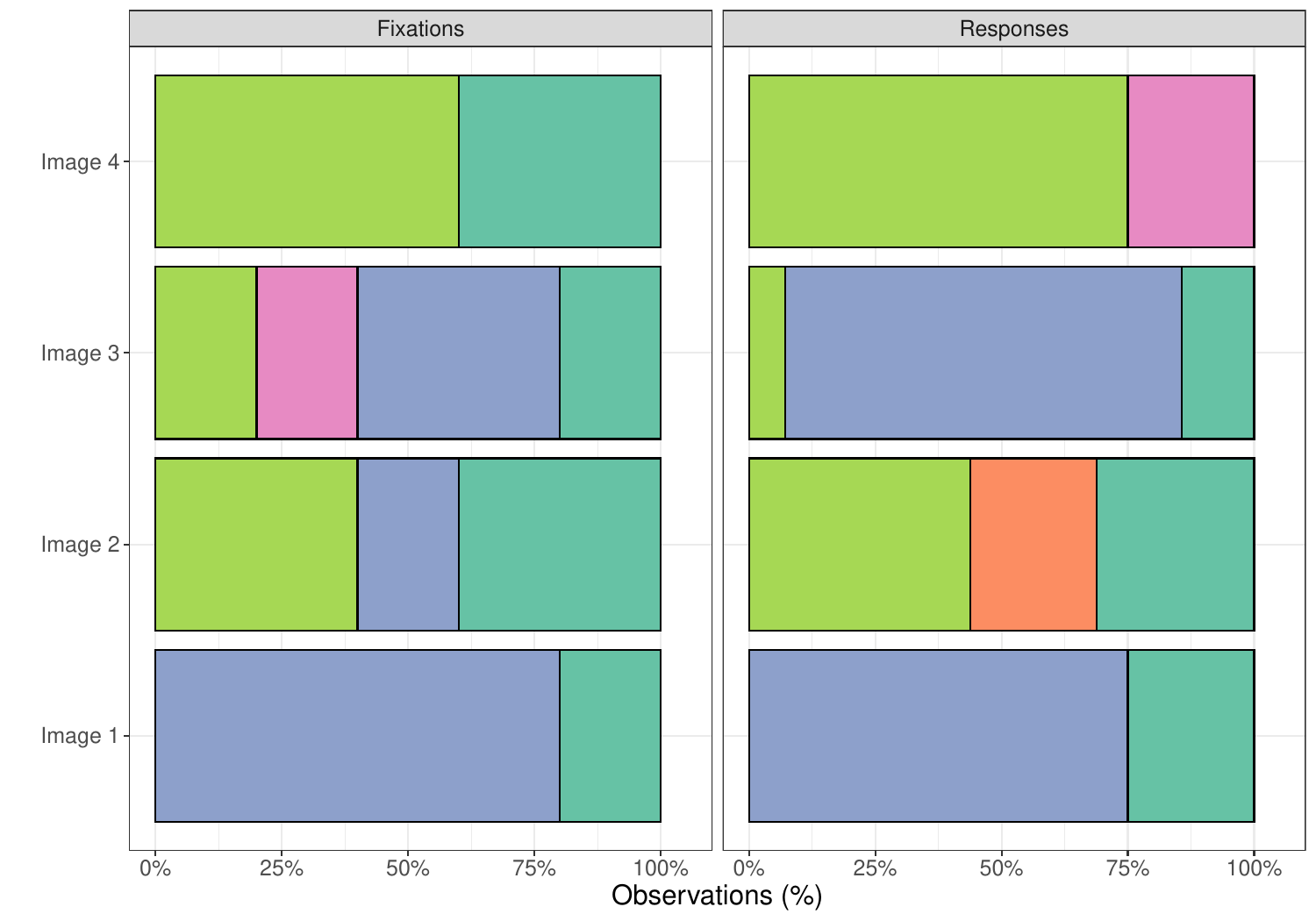}     
\end{subfigure}
\begin{subfigure}{.49\textwidth}
\caption{Gambles, longest dwell time}
    \includegraphics[width=0.95\textwidth]{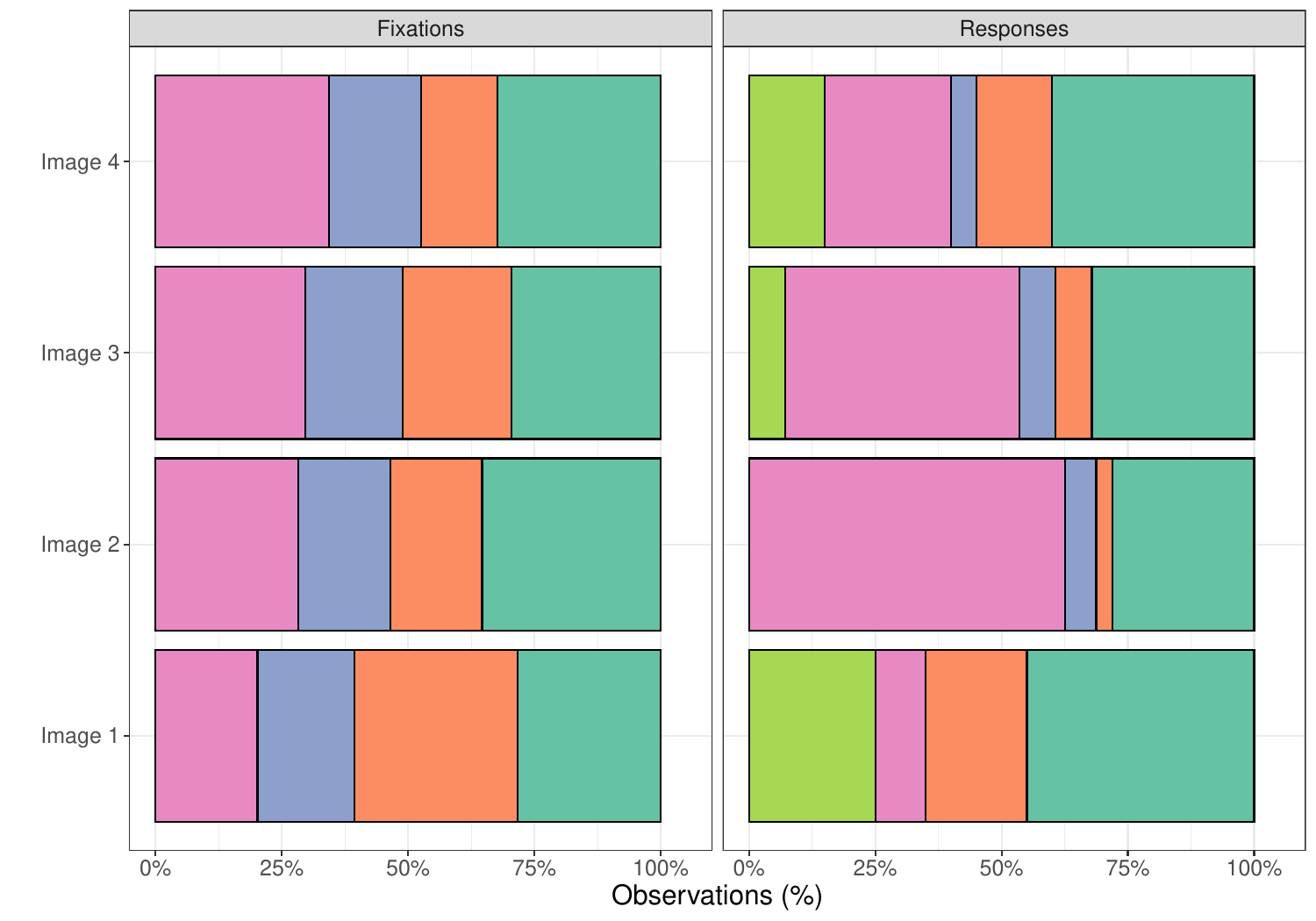}     
\end{subfigure}
\caption{Comparison of the distribution of fixations or dwell times across images (left panels) and the frequency of responses (right panels). First fixations and dwell times are both expressed as percentages (of all first fixations or the total dwell times). Color names have been omitted from the plots, taking up too much space, and not providing additional information.}
\label{fig:alternative_answers}
\end{figure*}

To examine whether such differences between the top answer and the second best answer plays a role Figure~\ref{fig:alternative_answers} examines how much each of the possible answers differ from each other. It plots the numbers of first fixations per AOI (left subplots; left subpanel) or the dwell times per AOI (right subplots; left subpanel). Both are expressed as a percentage (of the number of fixations, or the overall dwell time). Next to these plots the frequency of the various responses are shown (right subpanels of each plot). Only the `first' and `longest' responses are examined here, because the large number of response alternatives for `transitions' (combinations of two regions) makes plotting difficult. These plots combine responses across visualisations (scanpath, chord diagram, space-time cube, and scarfplot), because we are primarily interested in the link between competition between response alternatives and given responses in this analysis. 

For the first fixation, participants often picked the correct answer (also seen in the accuracy data). Incorrect responses were sometimes the second-best response (image 2, adverts; image 1, gambles), but sometimes also a different response (image 1, adverts; image 4 adverts; image 5 gambles). For the longest dwell times, accuracy was lower overall, and sometimes the second best answer was chosen (image 1, image 2, image 3 adverts). For the gambles the fixation data makes it clear that this task was much more difficult (the eye movement data are much more similar across AOIs).

\subsection{Participant features}

\begin{figure*}[h!]
\begin{subfigure}{.49\textwidth}
\caption{Education}
    \includegraphics[width=0.95\textwidth]{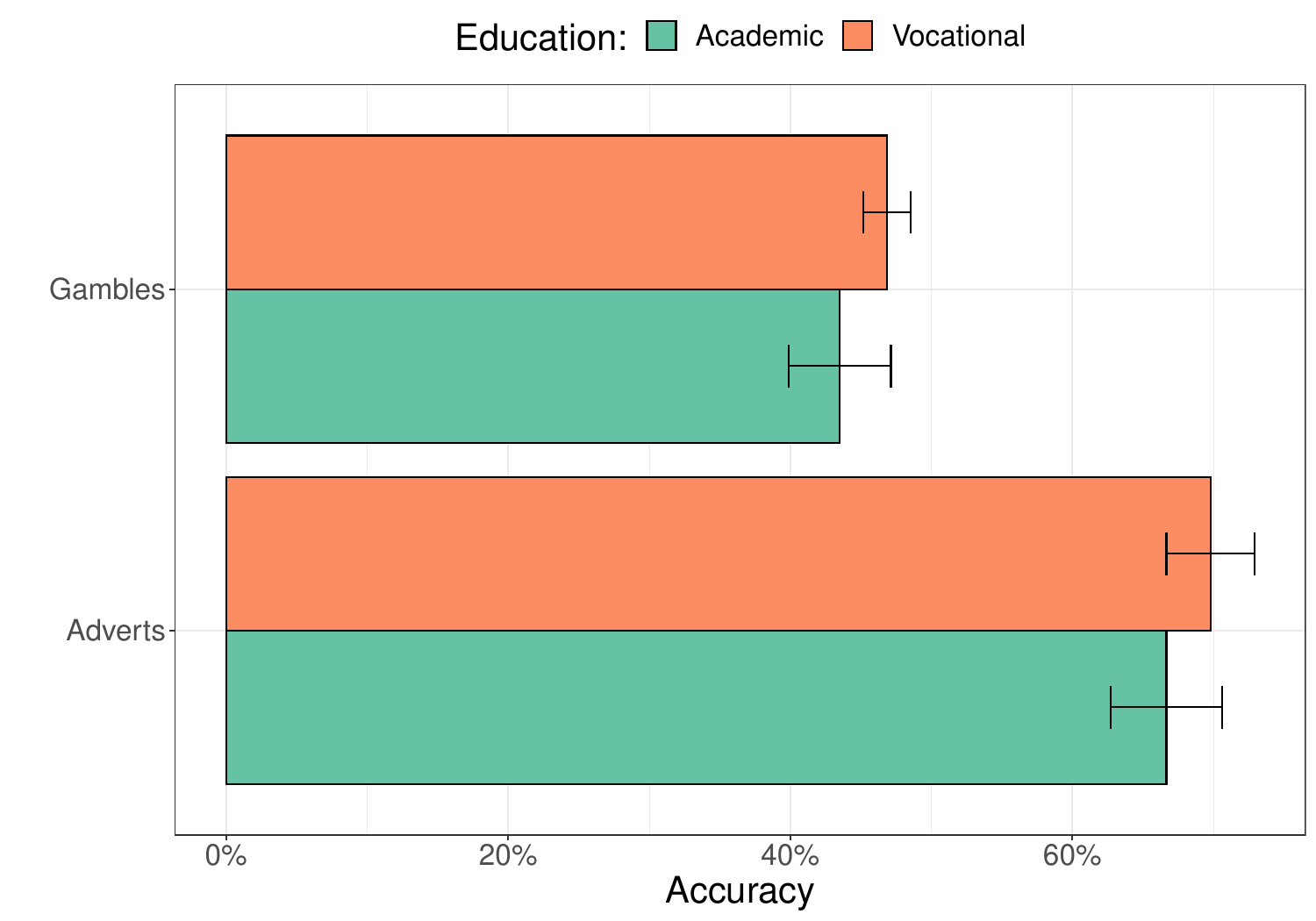}   
\end{subfigure}
\begin{subfigure}{.49\textwidth}
\caption{Graphs at work}
    \includegraphics[width=0.95\textwidth]{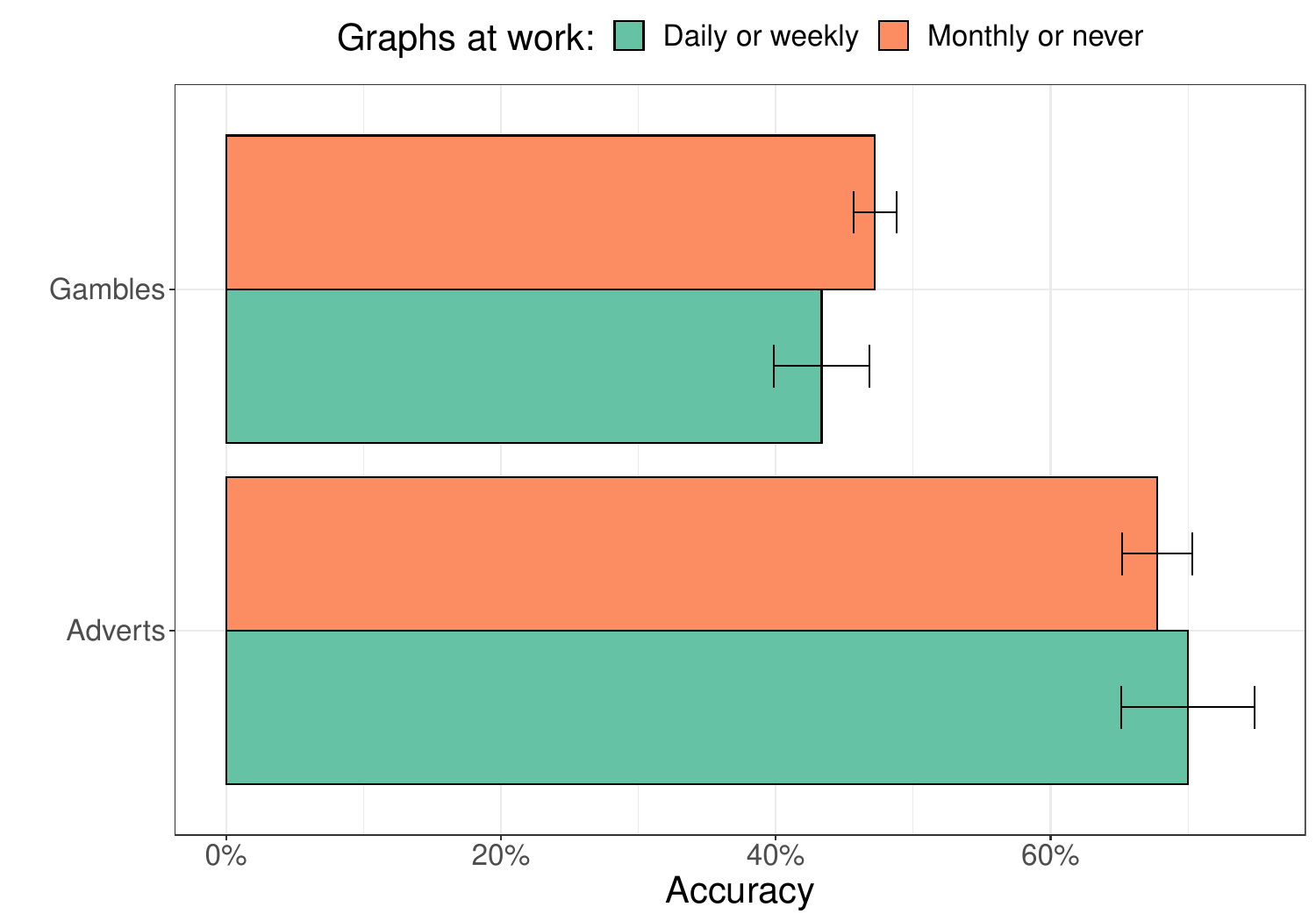}     
\end{subfigure}
\begin{subfigure}{.49\textwidth}
\caption{Graphs outside work}
    \includegraphics[width=0.95\textwidth]{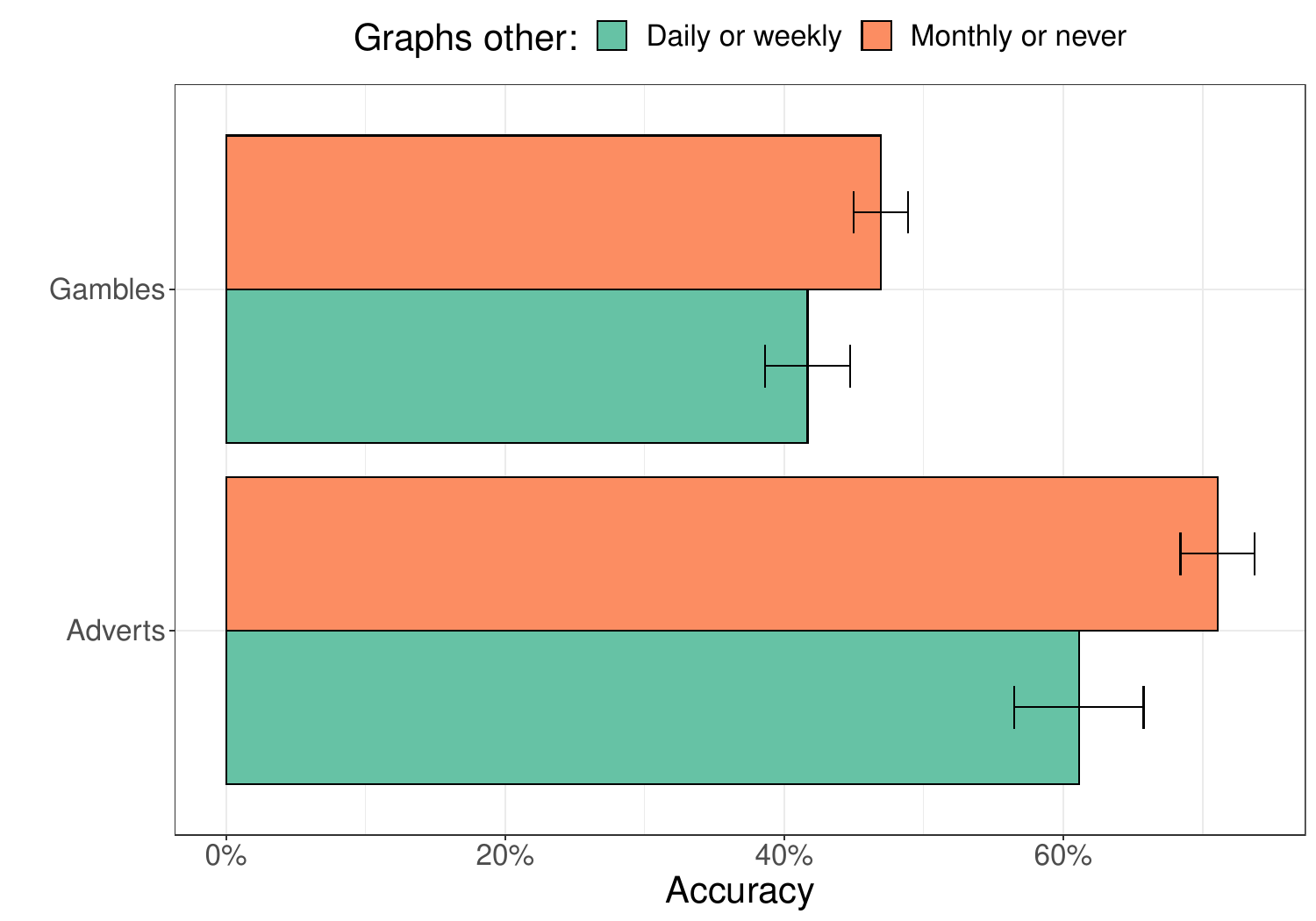}     
\end{subfigure}
\caption{Comparison of overall accuracy of participants with an a) academic or higher vocational training, b) participants who used graphs frequently or not frequently in a work setting and c) participants who used graphs frequently or not frequently in a non-work setting. Errorbars show the standard error of the mean across participants}
\label{fig:participants}
\end{figure*}

We asked participants to indicate the highest level of education that they attained as well as how often they used graphs in a work context or outside work. Figure~\ref{fig:participants} compares the overall accuracy of participants, based on their education, and reported use of graphs. We divided the data in two groups in such a way that there were similar numbers of participants in each group (although somewhat larger groups were obtained for vocational higher education, N = 16, and not frequently using graphs outside work, N = 19). Mixed effects logistic regressions did not show an interaction with the data-set for education ($\chi^2$(1) $<$ 0.001, $p$ = 0.98), work experience with graphs ($\chi^2$(1) = 0.56, $p$ = 0.45), and outside work experience with graphs ($\chi^2$(1) = 0.35, $p$ = 0.56). No main effects were found of education ($\chi^2$(1) = 0.63, $p$ = 0.43), work experience with graphs ($\chi^2$(1) = 0.043, $p$ = 0.84), and outside work experience with graphs ($\chi^2$(1) = 2.72, $p$ = 0.099), suggesting that education level and experience did not affect accuracy of the responses. This could be because the group was fairly homogeneous (no large differences in education or experiences).

\section{Discussion}

Eye movements contain both a spatial and a temporal component, which can be difficult to visualise, particularly when data from a large number of participants is presented \citep{rodrigues2018multiscale,peysakhovich2017scanpath}. User studies of the adequacy of different visualisations have often used a single data-set \citep[e.g.,][]{eraslan2016eye,menges2020visualization} and often focus on the user experience \citep{blascheck2017visualization,menges2020visualization}, instead on the accuracy of responses to questions about the data. We therefore ran a user study comparing four common spatio-temporal visualisations of eye tracking data, asking users three different questions about each visualisation and using two different data-sets (with different levels of expected clutter). Our results show that the accuracy of the responses to the questions depends on the combination of the visualisation, the type of question asked, and the type of data shown. Reported experience with using graphs or level of education did not affect the accuracy of the responses.

Overall, performance was lower for the scanpath, compared to the other visualisations. This may be due to the clutter often found in such graphs \citep{peysakhovich2017scanpath}, which is less of a problem in a 3D version (the space-time cube) that can be rotated. Another reason for the lower performance for the scanpath could be that it was the only type of graph where participants did not see the regions of interest directly. Instead, they needed to combine the scanpath image with the AOI image to determine their answer. The space-time cube did not require this additional processing step, because the AOI image was shown in the bottom of the cube, and the fixations were shown in the colour of their corresponding AOI. To determine whether this particular feature of the study design lowered accuracy for the scanpath, future studies should therefore study scanpaths that similarly make use of an overlay of the AOI image, or colours of fixations that represent the AOI.

The finding that the optimal visualisation depends on the task is in line with \cite{burch2021power}, who found that participants chose different combinations of visualisations when solving different tasks with an eye tracking data-set. In their study, participants preferred and most frequently selected the density plot (similar to a heatmap) and the bee swarm, compared to the scarfplot and the scanpath. Only when asked to compare the gaze sequence of three people using AOIs, the scarfplot was often preferred. In line with our findings, the scanpath was the least preferred visualisation. A direct comparison between our results and  \cite{burch2021power} is difficult, because the paper does provide data on how often the answers were correct and no data are comparable to our  `transitions' task. 

The importance of the task for the optimal visualisation is also described by \cite{kurzhals2015task}, who make a distinction of tasks between `when', `where' and `who' (the latter involving a comparison of viewers). A further distinction is made between `compare', `relate', and `detect' \citep{kurzhals2015task}. For tasks that focus on the `where' aspect of eye movements, the space-time cube is recommended (over the heatmap), as it also provides a temporal component. For questions about `when' participants look and `who' look, the scarfplot is suggested. For `compare' questions, adjacent scanpaths are suggested, whereas for `relate' questions, transition graphs are recommended. AOI rivers are suggested for `detect' questions. Our data suggest that the scarfplot may not be optimal for all `where' questions: it works well for questions about where participants look first (or last), but is less optimal for where participants look most. Our data confirm that a transition graph (we used a chord diagram) is best for `relate' questions.

Some of the present findings were anticipated from the nature of the questions and the visualisations, such as better performance on questions about eye movements between regions for the chord diagram (as this is a graph specifically designed to show transitions between states). We also anticipated higher levels of clutter in scanpaths when observers looked at the gambles, compared to the adverts, and therefore lower performance on the scanpaths for the gambles data-set. We decided to still apply a full design of four visualisations, three questions and two data-sets in order to obtain quantitative data on how well participants could perform the task for each combination, confirming our common sense hypotheses about which visualisation leads to better performance for which types of question. Importantly, our study provides quantification of the differences between graphs that is difficult to derive from common sense alone.

We found some variation in accuracy (Figure~\ref{fig:groups}) for the different images from the eye tracking experiment (Figure~\ref{fig:images}), meaning that answers were easier to extract from the visualisations for certain types of images. In an analysis that compared how strongly the top answer stood out and the frequency of answers in the survey, we found that some of the incorrect responses could be explained from close calls between the top answer and the second best answer. The variation in accuracy per original image indicates that it is important not to rely on a single stimulus in a user study.

Our study had some limitations that may be addressed in future studies. First, we only plotted data of five participants of the original eye tracking study. This is likely to have reduced the visual clutter of some of the visualisations, and the scanpath in particular. Still, we found that for the eye movements during the presentation of the two gambles that scanpaths were highly cluttered and participants were struggling to provide accurate answers to the questions. It is important to examine the role of the number of observers in  visualisations of eye movements, as it may mean that interactive visualisations may be needed, where a subset of observers can be selected. Typically, eye movement studies have around 20 to 30 participants \citep[e.g.,][]{birmingham2008social,crosby2019does}, meaning that if problems with clutter cannot be resolved in visualisations, selection of observers in interactive versions may be required. Alternative approaches could involve the clustering of scanpaths to reduce clutter \citep[e.g.,][]{eraslan2016eye,rodrigues2018multiscale,li2017scanpath}. While such methods reduce clutter, they will also obscure variations between participants, and results will depend on how scanpaths are clustered. 

We also had to restrict our comparison to four possible visualisations, three questions, and two types of eye tracking data. We chose three fairly common questions that can be asked about eye movement data (where people look first, the longest and how often they look between regions). These are fairly general questions, meaning that we could not evaluate the fit of the various visualisations for more specific questions. 

Our visualisations did not display any group information of participants in the eye tracking data-set. Often eye tracking studies compare conditions (e.g., people viewing low and high value objects in adverts) or groups (e.g., older and younger participants). The present visualisations may be less suited for such group comparisons, because they focus on the individual participants. Group comparisons, for example on dwell times, are often presented in the form of bar graphs with errorbars or boxplots, because these allow for an easier comparison of the groups or conditions than more complex plots such as the scanpaths and scarfplots \citep{blascheck2017visualization}. Future research should determine whether such summary visualisations (e.g., bar graph or boxplots) are indeed better for comparing groups than the visualisations used here (e.g., scarfplots in different subplots for each group) using a similar approach as used in the present study. 

We chose one specific version of a graph to display how often observers looked between regions (the chord diagram). Various other visualisations have been proposed for this purpose, such as the transition matrix \citep{goldberg1999computer} or a directed graph \citep{blascheck2017visualization}. Future research should examine whether any of these alternatives to the chord diagram lead to better understanding of the data, and whether such diagrams can be extended to show longer sequences than the combinations of two subsequent fixations (e.g., which trigrams of AOIs are commonly found).

While the eye tracking data-sets that we used (freely viewing adverts, forced choice between gambles) are fairly typical for eye tracking studies, both data-sets involve static images. Future studies should therefore look into methods to visualise eye movements for movie clips and animations \citep[e.g.,][]{kuhn2005magic}, human computer interaction situations \citep{morimoto2005eye} and real-world applications \citep[e.g.,][]{land1996relations,land2001ways,foulsham2014top}. 

Future studies should also address the question whether some visualisations are better when the measurement interval is variable, rather than fixed. For our adverts eye tracking data, a fixed measurement interval was used. For our gambles data, the measurement interval depended on when participants pressed a key to indicate their decision. When the measurement interval becomes highly variable between participants, the four visualisations in our study may provide less insight: bars in the scarfplot will vary in length, as will the lines in the scanpaths and space-time cubes. The chord diagram may highlight the eye movement patterns of participants with longer measurement intervals more strongly. Specific visualisations methods may therefore be required that can deal with strong variations in recording intervals per participant.

Our study used a sample of 25 participants. Inspection of the data and the variability of the results indicate that this number sufficed to answer the main question whether the optimal visualisation depends on the type of question and the data-set. For some comparisons a larger sample may be needed, as well as a more diverse sample. We did not find an effect of the expertise of our participants. This could be due to the rather homogeneous nature of opportunity sample that we used (leading to mostly people with a BSc or MSc degree, who had similar experience with graphs). Furthermore, we measured graph experience by asking how often participants worked with graphs. More standardised methods to measure graph literacy would be advisable in future studies \citep{garcia2016measuring,okan2019using}.

Finally, the current visualisations mostly focused on general data exploration. Studies increasingly make use of machine learning and deep learning methods, for example to predict choice from patterns of eye movements \citep{gere2021predict,unger2023predicting}. Future research should therefore establish which types of visualisations are most effective for exploratory data analysis in this context.

\subsection*{Conclusion}

Various types of visualisations have been proposed for eye tracking data. The present study suggests that the optimal visualisation depends on the question about the eye tracking data and the data-set. When presenting eye movement data it may therefore be beneficial to jointly present multiple visualisations so that the reader can explore the data for multiple aspects.

\bibliographystyle{unsrtnat}
\bibliography{eye_movement_vis}

\begin{thebibliography}{45}
\providecommand{\natexlab}[1]{#1}
\providecommand{\url}[1]{\texttt{#1}}
\expandafter\ifx\csname urlstyle\endcsname\relax
  \providecommand{\doi}[1]{doi: #1}\else
  \providecommand{\doi}{doi: \begingroup \urlstyle{rm}\Url}\fi

\bibitem[Rayner(1978)]{rayner1978eye}
Keith Rayner.
\newblock Eye movements in reading and information processing.
\newblock \emph{Psychological bulletin}, 85\penalty0 (3):\penalty0 618, 1978.

\bibitem[Rayner(1998)]{rayner1998eye}
Keith Rayner.
\newblock Eye movements in reading and information processing: 20 years of
  research.
\newblock \emph{Psychological bulletin}, 124\penalty0 (3):\penalty0 372, 1998.

\bibitem[Rayner et~al.(2005)Rayner, Juhasz, and Pollatsek]{rayner2005eye}
Keith Rayner, Barbara~J Juhasz, and Alexander Pollatsek.
\newblock Eye movements during reading.
\newblock 2005.

\bibitem[Land and Horwood(1996)]{land1996relations}
Michael~F Land and Julia Horwood.
\newblock The relations between head and eye movements during driving.
\newblock \emph{Vision in vehicles}, 5:\penalty0 153--160, 1996.

\bibitem[Land and Hayhoe(2001)]{land2001ways}
Michael~F Land and Mary Hayhoe.
\newblock In what ways do eye movements contribute to everyday activities?
\newblock \emph{Vision research}, 41\penalty0 (25-26):\penalty0 3559--3565,
  2001.

\bibitem[Blascheck et~al.(2017)Blascheck, Kurzhals, Raschke, Burch, Weiskopf,
  and Ertl]{blascheck2017visualization}
Tanja Blascheck, Kuno Kurzhals, Michael Raschke, Michael Burch, Daniel
  Weiskopf, and Thomas Ertl.
\newblock Visualization of eye tracking data: A taxonomy and survey.
\newblock In \emph{Computer Graphics Forum}, volume~36, pages 260--284. Wiley
  Online Library, 2017.

\bibitem[Bojko(2009)]{bojko2009informative}
Agnieszka~Aga Bojko.
\newblock Informative or misleading? heatmaps deconstructed.
\newblock In \emph{International conference on human-computer interaction},
  pages 30--39. Springer, 2009.

\bibitem[Menges et~al.(2020)Menges, Kramer, Hill, Nisslmueller, Kumar, and
  Staab]{menges2020visualization}
Raphael Menges, Sophia Kramer, Stefan Hill, Marius Nisslmueller, Chandan Kumar,
  and Steffen Staab.
\newblock A visualization tool for eye tracking data analysis in the web.
\newblock In \emph{ACM Symposium on Eye Tracking Research and Applications},
  pages 1--5, 2020.

\bibitem[Burch et~al.(2021)Burch, Wallner, Broeks, Piree, Boonstra, Vlaswinkel,
  Franken, and Van~Wijk]{burch2021power}
Michael Burch, G{\"u}nter Wallner, Nick Broeks, Lulof Piree, Nynke Boonstra,
  Paul Vlaswinkel, Silke Franken, and Vince Van~Wijk.
\newblock The power of linked eye movement data visualizations.
\newblock In \emph{ACM Symposium on Eye Tracking Research and Applications},
  pages 1--11, 2021.

\bibitem[Bakardzhiev et~al.(2021)Bakardzhiev, Burgt, Martins, Dool, Jansen,
  Scheppingen, Wallner, and Burch]{bakardzhiev2021web}
Hristo Bakardzhiev, Marloes van~der Burgt, Eduardo Martins, Bart van~den Dool,
  Chyara Jansen, David~van Scheppingen, G{\"u}nter Wallner, and Michael Burch.
\newblock A web-based eye tracking data visualization tool.
\newblock In \emph{International Conference on Pattern Recognition}, pages
  405--419. Springer, 2021.

\bibitem[Noton and Stark(1971{\natexlab{a}})]{noton1971scanpaths_a}
David Noton and Lawrence Stark.
\newblock Scanpaths in eye movements during pattern perception.
\newblock \emph{Science}, 171\penalty0 (3968):\penalty0 308--311,
  1971{\natexlab{a}}.

\bibitem[Peysakhovich and Hurter(2017)]{peysakhovich2017scanpath}
Vsevolod Peysakhovich and Christophe Hurter.
\newblock Scanpath visualization and comparison using visual aggregation
  techniques.
\newblock \emph{Journal of Eye Movement Research}, 10\penalty0 (5), 2017.

\bibitem[Rodrigues et~al.(2018)Rodrigues, Netzel, Spalink, and
  Weiskopf]{rodrigues2018multiscale}
Nils Rodrigues, Rudolf Netzel, Joachim Spalink, and Daniel Weiskopf.
\newblock Multiscale scanpath visualization and filtering.
\newblock In \emph{Proceedings of the 3rd Workshop on Eye Tracking and
  Visualization}, pages 1--5, 2018.

\bibitem[Eraslan et~al.(2016)Eraslan, Yesilada, and Harper]{eraslan2016eye}
Sukru Eraslan, Yeliz Yesilada, and Simon Harper.
\newblock Eye tracking scanpath analysis techniques on web pages: A survey,
  evaluation and comparison.
\newblock \emph{Journal of Eye Movement Research}, 9\penalty0 (1), 2016.

\bibitem[Kurzhals et~al.(2015{\natexlab{a}})Kurzhals, Hlawatsch, Heimerl,
  Burch, Ertl, and Weiskopf]{kurzhals2015gaze}
Kuno Kurzhals, Marcel Hlawatsch, Florian Heimerl, Michael Burch, Thomas Ertl,
  and Daniel Weiskopf.
\newblock Gaze stripes: Image-based visualization of eye tracking data.
\newblock \emph{IEEE transactions on visualization and computer graphics},
  22\penalty0 (1):\penalty0 1005--1014, 2015{\natexlab{a}}.

\bibitem[Blascheck et~al.(2014)Blascheck, Kurzhals, Raschke, Burch, Weiskopf,
  and Ertl]{blascheck2014state}
Tanja Blascheck, Kuno Kurzhals, Michael Raschke, Michael Burch, Daniel
  Weiskopf, and Thomas Ertl.
\newblock State-of-the-art of visualization for eye tracking data.
\newblock In \emph{EuroVis (STARs)}, 2014.

\bibitem[Noton and Stark(1971{\natexlab{b}})]{noton1971scanpaths}
David Noton and Lawrence Stark.
\newblock Scanpaths in saccadic eye movements while viewing and recognizing
  patterns.
\newblock \emph{Vision research}, 11\penalty0 (9):\penalty0 929--IN8,
  1971{\natexlab{b}}.

\bibitem[Hembrooke et~al.(2006)Hembrooke, Feusner, and
  Gay]{hembrooke2006averaging}
Helene Hembrooke, Matt Feusner, and Geri Gay.
\newblock Averaging scan patterns and what they can tell us.
\newblock In \emph{Proceedings of the 2006 symposium on eye tracking research
  \& applications}, pages 41--41, 2006.

\bibitem[Chen et~al.(2013)Chen, Alves, and Sol]{chen2013combining}
Monchu Chen, Nelson Alves, and Ricardo Sol.
\newblock Combining spatial and temporal information of eye movements in
  goal-oriented tasks.
\newblock In \emph{International Conference on Human Factors in Computing and
  Informatics}, pages 827--830. Springer, 2013.

\bibitem[Hurter et~al.(2013)Hurter, Ersoy, Fabrikant, Klein, and
  Telea]{hurter2013bundled}
Christophe Hurter, Ozan Ersoy, Sara~Irina Fabrikant, Tijmen~R Klein, and
  Alexandru~C Telea.
\newblock Bundled visualization of dynamicgraph and trail data.
\newblock \emph{IEEE transactions on visualization and computer graphics},
  20\penalty0 (8):\penalty0 1141--1157, 2013.

\bibitem[Elmqvist(2005)]{elmqvist2005balloonprobe}
Niklas Elmqvist.
\newblock Balloonprobe: Reducing occlusion in 3d using interactive space
  distortion.
\newblock In \emph{Proceedings of the ACM symposium on Virtual reality software
  and technology}, pages 134--137, 2005.

\bibitem[Elmqvist and Tsigas(2008)]{elmqvist2008taxonomy}
Niklas Elmqvist and Philippas Tsigas.
\newblock A taxonomy of 3d occlusion management for visualization.
\newblock \emph{IEEE transactions on visualization and computer graphics},
  14\penalty0 (5):\penalty0 1095--1109, 2008.

\bibitem[Kurzhals and Weiskopf(2013)]{kurzhals2013space}
Kuno Kurzhals and Daniel Weiskopf.
\newblock Space-time visual analytics of eye-tracking data for dynamic stimuli.
\newblock \emph{IEEE Transactions on Visualization and Computer Graphics},
  19\penalty0 (12):\penalty0 2129--2138, 2013.

\bibitem[Kurzhals et~al.(2015{\natexlab{b}})Kurzhals, Burch, Blascheck,
  Andrienko, Andrienko, and Weiskopf]{kurzhals2015task}
Kuno Kurzhals, Michael Burch, Tanja Blascheck, Gennady Andrienko, Natalia
  Andrienko, and Daniel Weiskopf.
\newblock A task-based view on the visual analysis of eye-tracking data.
\newblock In \emph{Workshop on Eye Tracking and Visualization}, pages 3--22.
  Springer, 2015{\natexlab{b}}.

\bibitem[Richardson and Dale(2005)]{richardson2005looking}
Daniel~C Richardson and Rick Dale.
\newblock Looking to understand: The coupling between speakers' and listeners'
  eye movements and its relationship to discourse comprehension.
\newblock \emph{Cognitive science}, 29\penalty0 (6):\penalty0 1045--1060, 2005.

\bibitem[Yang and Wacharamanotham(2018)]{yang2018alpscarf}
Chia-Kai Yang and Chat Wacharamanotham.
\newblock Alpscarf: Augmenting scarf plots for exploring temporal gaze
  patterns.
\newblock In \emph{Extended abstracts of the 2018 CHI conference on human
  factors in computing systems}, pages 1--6, 2018.

\bibitem[Goldberg and Kotval(1999)]{goldberg1999computer}
Joseph~H Goldberg and Xerxes~P Kotval.
\newblock Computer interface evaluation using eye movements: methods and
  constructs.
\newblock \emph{International journal of industrial ergonomics}, 24\penalty0
  (6):\penalty0 631--645, 1999.

\bibitem[Krejtz et~al.(2013)Krejtz, Szarkowska, and Krejtz]{krejtz2013effects}
Izabela Krejtz, Agnieszka Szarkowska, and Krzysztof Krejtz.
\newblock The effects of shot changes on eye movements in subtitling.
\newblock 2013.

\bibitem[Blascheck et~al.(2013)Blascheck, Raschke, and
  Ertl]{blascheck2013circular}
Tanja Blascheck, Michael Raschke, and Thomas Ertl.
\newblock Circular heat map transition diagram.
\newblock In \emph{Proceedings of the 2013 Conference on Eye Tracking South
  Africa}, pages 58--61, 2013.

\bibitem[Goldberg and Helfman(2011)]{goldberg2011eye}
Joseph~H Goldberg and Jonathan Helfman.
\newblock Eye tracking for visualization evaluation: Reading values on linear
  versus radial graphs.
\newblock \emph{Information visualization}, 10\penalty0 (3):\penalty0 182--195,
  2011.

\bibitem[Rees et~al.(2020)Rees, Laramee, Brookes, and
  D'Cruze]{rees2020interaction}
Dylan Rees, Robert~S Laramee, Paul Brookes, and Tony D'Cruze.
\newblock Interaction techniques for chord diagrams.
\newblock In \emph{2020 24th International Conference Information Visualisation
  (IV)}, pages 28--37, 2020.

\bibitem[Wang(2021)]{wang2021eye}
Zhiguo Wang.
\newblock Eye movement data analysis and visualization.
\newblock In \emph{Eye-Tracking with Python and Pylink}, pages 197--224.
  Springer, 2021.

\bibitem[R{\"a}ih{\"a} et~al.(2005)R{\"a}ih{\"a}, Aula, Majaranta, Rantala, and
  Koivunen]{raiha2005static}
Kari-Jouko R{\"a}ih{\"a}, Anne Aula, P{\"a}ivi Majaranta, Harri Rantala, and
  Kimmo Koivunen.
\newblock Static visualization of temporal eye-tracking data.
\newblock In \emph{IFIP Conference on Human-Computer Interaction}, pages
  946--949. Springer, 2005.

\bibitem[Burch et~al.(2013)Burch, Kull, and Weiskopf]{burch2013aoi}
Michael Burch, Andreas Kull, and Daniel Weiskopf.
\newblock Aoi rivers for visualizing dynamic eye gaze frequencies.
\newblock In \emph{Computer Graphics Forum}, volume~32, pages 281--290. Wiley
  Online Library, 2013.

\bibitem[Riehmann et~al.(2005)Riehmann, Hanfler, and
  Froehlich]{riehmann2005interactive}
Patrick Riehmann, Manfred Hanfler, and Bernd Froehlich.
\newblock Interactive sankey diagrams.
\newblock In \emph{IEEE Symposium on Information Visualization, 2005. INFOVIS
  2005.}, pages 233--240. IEEE, 2005.

\bibitem[Birmingham et~al.(2008)Birmingham, Bischof, and
  Kingstone]{birmingham2008social}
Elina Birmingham, Walter~F Bischof, and Alan Kingstone.
\newblock Social attention and real-world scenes: The roles of action,
  competition and social content.
\newblock \emph{Quarterly journal of experimental psychology}, 61\penalty0
  (7):\penalty0 986--998, 2008.

\bibitem[Crosby and Hermens(2019)]{crosby2019does}
Freya Crosby and Frouke Hermens.
\newblock Does it look safe? an eye tracking study into the visual aspects of
  fear of crime.
\newblock \emph{Quarterly Journal of Experimental Psychology}, 72\penalty0
  (3):\penalty0 599--615, 2019.

\bibitem[Li et~al.(2017)Li, Zhang, and Chen]{li2017scanpath}
Aoqi Li, Yingxue Zhang, and Zhenzhong Chen.
\newblock Scanpath mining of eye movement trajectories for visual attention
  analysis.
\newblock In \emph{2017 IEEE International Conference on Multimedia and Expo
  (ICME)}, pages 535--540. IEEE, 2017.

\bibitem[Kuhn and Tatler(2005)]{kuhn2005magic}
Gustav Kuhn and Benjamin~W Tatler.
\newblock Magic and fixation: Now you don't see it, now you do.
\newblock \emph{Perception}, 34\penalty0 (9):\penalty0 1155--1161, 2005.

\bibitem[Morimoto and Mimica(2005)]{morimoto2005eye}
Carlos~H Morimoto and Marcio~RM Mimica.
\newblock Eye gaze tracking techniques for interactive applications.
\newblock \emph{Computer vision and image understanding}, 98\penalty0
  (1):\penalty0 4--24, 2005.

\bibitem[Foulsham et~al.(2014)Foulsham, Chapman, Nasiopoulos, and
  Kingstone]{foulsham2014top}
Tom Foulsham, Craig Chapman, Eleni Nasiopoulos, and Alan Kingstone.
\newblock Top-down and bottom-up aspects of active search in a real-world
  environment.
\newblock \emph{Canadian Journal of Experimental Psychology/Revue canadienne de
  psychologie exp{\'e}rimentale}, 68\penalty0 (1):\penalty0 8, 2014.

\bibitem[Garcia-Retamero et~al.(2016)Garcia-Retamero, Cokely, Ghazal, and
  Joeris]{garcia2016measuring}
Rocio Garcia-Retamero, Edward~T Cokely, Saima Ghazal, and Alexander Joeris.
\newblock Measuring graph literacy without a test: a brief subjective
  assessment.
\newblock \emph{Medical Decision Making}, 36\penalty0 (7):\penalty0 854--867,
  2016.

\bibitem[Okan et~al.(2019)Okan, Janssen, Galesic, and Waters]{okan2019using}
Yasmina Okan, Eva Janssen, Mirta Galesic, and Erika~A Waters.
\newblock Using the short graph literacy scale to predict precursors of health
  behavior change.
\newblock \emph{Medical Decision Making}, 39\penalty0 (3):\penalty0 183--195,
  2019.

\bibitem[Gere et~al.(2021)Gere, H{\'e}berger, and Kov{\'a}cs]{gere2021predict}
Attila Gere, K{\'a}roly H{\'e}berger, and S{\'a}ndor Kov{\'a}cs.
\newblock How to predict choice using eye-movements data?
\newblock \emph{Food Research International}, 143:\penalty0 110309, 2021.

\bibitem[Unger et~al.(2023)Unger, Wedel, and Tuzhilin]{unger2023predicting}
Moshe Unger, Michel Wedel, and Alexander Tuzhilin.
\newblock Predicting consumer choice from raw eye-movement data using the
  retina deep learning architecture.
\newblock \emph{Available at SSRN 4341410}, 2023.

\end{thebibliography}
\end{document}